\documentclass{ifacconf}

\usepackage{graphicx}      
\usepackage{natbib}        
\usepackage{amsmath}

\usepackage{lipsum}
\newcommand\blfootnote[1]{%
	\begingroup
	\renewcommand\thefootnote{}\footnote{#1}%
	\addtocounter{footnote}{-1}%
	\endgroup
}

\begin{document}
\begin{frontmatter}

\title{\LARGE \bf
Identification of MISO systems in Minimal Realization Form
}

\author[]{Chaithanya K. Donda,}
\author[third]{Deepak Maurya,} 
\author[third]{Arun K. Tangirala,}
\author[third]{Shankar Narasimhan}

\address[third]{Systems \& Controls Group, Indian Institute of Technology Madras, Chennai (e-mail: chaitu.7.d@gmail.com, maurya@cse.iitm.ac.in, arunkt@iitm.ac.in, naras@iitm.ac.in)}

\begin{abstract}                
The paper is concerned with identifying transfer functions of individual input channels in minimal realization form of a Multi-Input Single Output (MISO) from the input-output data corrupted by the error in all the variables. Such a framework is commonly referred to as error-in-variables (EIV). A common approach in the existing methods for identification of MISO systems is to estimate a non-minimal order transfer function under a subset of simplistic assumptions like homoskedastic error variances, known order, and delay. In this work, we deal with the challenging problem of identifying order, delay in each input of minimal realization form separately while estimating the transfer functions. 
We also estimate the heteroskedastic noise variances in each of the multiple inputs and output variables. An automated approach for the identification of MISO systems of minimal realization form in the EIV framework is proposed. Numerical case studies are presented to illustrate the efficacy of the proposed algorithm in identifying the transfer function along with the order, delay, and noise variances.
\end{abstract}


\end{frontmatter}

\section{Introduction}
\begin{figure}[thpb]
  \centering
  \includegraphics[scale=0.45]{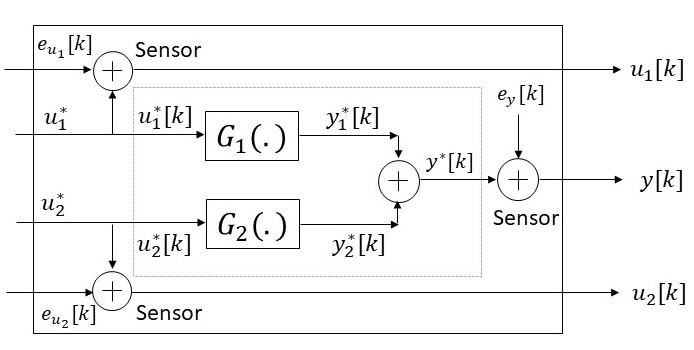}
  \caption{Linear Dynamic EIV Architecture for a two-input single-output system}
  \label{fig:miso_eiv}
\end{figure}

\blfootnote{\textsuperscript{\textcopyright} 20XX the authors. This work has been accepted to IFAC for publication under a Creative Commons Licence CC-BY-NC-ND}
Model identification of multi-input multi-output (MIMO) systems is a crucial problem in the field of process control and has various applications. The identification of such systems can be simplified to the identification of multi-input single-output (MISO) models for each open-loop system. In general, both the inputs and outputs may be measured with errors. In such a case, the problem is commonly referred to as error-in-variables (EIV) model identification. A typical MISO system with two inputs is shown in Figure \ref{fig:miso_eiv}. It can be noticed that the noise-free inputs $u_1^{\star}$, $u_2^{\star}$ and noise-free output $y^{\star}$ is corrupted with noise variables $e_{u_1}$, $e_{u_2}$ and $e_y$ respectively. 
One of the challenging aspect of this problem is identification of transfer function $G_1$ and $G_2$ from the measured noisy variables $u_1$, $u_2$ and $y$. In this work, we do not assume that the input-output orders and delays are available. Another key aspect is that we intend to identify the minimal order transfer function. This phenomenon is explained below. 

Let the transfer functions be denoted by $G_1(q^{-1}) = \frac{B_1(q^{-1})}{A_1(q^{-1})}$ and $G_2(q^{-1}) = \frac{B_2(q^{-1})}{A_2(q^{-1})}$, where $q^{-1}$ is the usual delay operator. With slight abuse of terminology, the output can be expressed as:
{\small 
\begin{align}
    y[k] &= \frac{B_1(q^{-1})}{A_1(q^{-1})} u_1[k] + \frac{B_2(q^{-1})}{A_2(q^{-1})} u_2[k] \label{eq:minimal_form} \\
    &= \frac{B_1(q^{-1}) A_2(q^{-1}) }{A_1(q^{-1}) A_2(q^{-1}) } u_1[k] + \frac{B_2(q^{-1}) A_1(q^{-1})}{A_1(q^{-1}) A_2(q^{-1})} u_2[k] \label{eq:non_minform}
\end{align}
}
The system as described in Eq. \eqref{eq:minimal_form} is in minimal realization form, but it can be equivalently expressed as shown in Eq. \eqref{eq:non_minform}. The later is termed as non-minimal form due to the excess order arising from the product terms like $A_1(q^{-1}) A_2(q^{-1})$. Most of the existing approaches estimate the transfer functions in this form for a given guess of delay and input-output order and further perform pole-zero cancellation to arrive at the minimal form. Nevertheless, this step is usually an approximation step based on heuristic approaches primarily due to two reasons. First is no exact pole-zero cancellation as the estimated numerical values are approximately equal but not exactly equal for the noisy data. Second is the original order of the transfer function $G_1$or $G_2$ in minimal form is unknown, which is an essential parameter required in pole-zero cancellation. In this paper, we propose a novel approach to resolve this issue, as illustrated later on.

Several works on MISO system identification can be seen in literature \citep{book:ljung2001system}. For example,  \cite{paper:ding2006bias} uses the bias compensation approach incorporated with the recursive least squares framework. Another interesting approach for improving the convergence rate is proposed by \cite{liu2009multi}. It utilizes the stochastic gradient algorithm using the multi-innovation theory. Applying DIPCA \cite{paper:dipca_dycops} to a MISO data would result in the higher-order realization of the transfer functions. Ideally, if the data is noise-free, there will be common poles and zeros, resulting in exact pole-zero cancellations and yield minimum realization of the transfer functions. However, when working with noisy data, the numerical estimates of poles and zeros will not be equal but will be very close to each other and do not result in exact pole-zero cancellation. 

This work is strongly motivated by dynamic iterative principal component analysis (DIPCA) algorithm \citep{paper:dipca_dycops} originally proposed for single input single output systems. We modify this framework to estimate the minimal realization transfer function. The key idea is to estimate the transfer function with respect to each input separately. The output is decomposed as the sum of the individual response from different inputs. While modeling each input separately, we also utilize the autocovariance function of decomposed output. This is done in an iterative manner using the Wiener-Khinchin theorem. The detailed description of the paper can be found in later sections. 

The rest of the paper is organized as follows. Section \ref{sec:found} briefly explains the basic ideas behind principal component analysis (PCA), iterative PCA, and dynamic iterative PCA in solving the EIV identification problem. In Section \ref{sec:prop_algo}, identification of the transfer functions in their minimum realization form, which is the main contribution of this work is discussed. In Section \ref{sec:siml_stud}, the simulation results are discussed, wherein Monte-Carlo simulations are presented to study the goodness of estimates of a two-input single-output system. The paper ends with a few concluding remarks in Section \ref{sec:conc}.

\section{FOUNDATIONS} 
\label{sec:found}
We begin with a discussion on the identification of linear static models using PCA. This will be followed up with an introduction to Iterative PCA (IPCA) for the same purpose but to solve a broader class of problems under generalized assumptions.

Let us consider a system containing $M$ variables related to each other linearly by \textit{d} equations, i.e., 
\begin{align}
    \mathbf{A}_0 \mathbf{X} = \mathbf{0} \label{eq:mdl_nf}
\end{align}
where $\mathbf{X} \in \mathbf{R}^{M \times N}$ is a collection of $N$ samples of $M$ variables and $\mathbf{A}_0 \in \mathbf{R}^{d \times M}$ denotes the constraint matrix or model or a basis of linear relations among $M$ variables. 
\begin{align}
    \mathbf{X} &= [\mathbf{x}[1],\mathbf{x}[2],...\mathbf{x}[N]]  \nonumber \\
 \text{where}  \qquad \mathbf{x}[k] &= [x_1[k],x_2[k],...,x_M[k]]^T
\end{align}
Here, $x_i[j]$ denotes the \textit{noise-free} measurement of the variable $x_i$ at $j^{th}$ instant and $\mathbf{x}[i]$ denotes noise-free measurement of $M$ variables at $i^{th}$ instant. 

The objective of the problem is to identify the constraint matrix $\mathbf{A}_0$ and the row dimension, \textit{d} from given $N$ samples of $M$ variables (which is $\mathbf{X}$). This can be easily solved by PCA \citep{jolliffe2011principal}, as discussed shortly. 

The EIV identification problem is to identify the constraints in Eq. \ref{eq:mdl_nf} from the noisy measurements of $\mathbf{X}$
\begin{align}
    \mathbf{z}[k] & = \mathbf{x}[k] + \mathbf{e}[k] \\
    \mathbf{Z} &= \mathbf{X} + \mathbf{E}
\end{align}
where $\mathbf{e}[k]$ is a white noise vector consisting errors with noise covariance $\mathbf{\Sigma_e}$ and $\mathbf{E}$ is a collection of $N$ samples.

In the following sections we briefly review  PCA and IPCA for identification problem in EIV framework.

\subsection{Principal Component Analysis (PCA)}
PCA \citep{jolliffe2011principal} is a popular multivariate statistical analysis tool that searches for correlation among the columns of a matrix through a search for zero eigenvalues of the covariance matrix. For the noise-free case, the number of zero eigenvalues of the sample covariance matrix denoted by $\mathbf{S_z} = \frac{1}{N}\mathbf{Z}^T \mathbf{Z}$ and corresponding eigenvectors provide the dimensionality and a basis for the constraint matrix, \textbf{A} respectively.

However, the constraint matrix estimate is unbiased only for homoskedastic errors case, meaning the variances of errors in all variables are equal and are spatially uncorrelated, i.e., $\mathbf{\Sigma_e} = \sigma^{2}_{e}\mathbf{I}_{M\times M}$. In such a scenario all the last \textit{d} eigenvalues are equal to $\sigma^{2}_{e}$ and constraint matrix, $\hat{\mathbf{A}}$ estimate is unbiased. This is a restrictive assumption that doesn't always hold. The general case, where the variances of errors in all variables are not equal, commonly referred to as \textit{heteroskedastic errors} case is handled by IPCA as described in the next sub-section.

\subsection{Iterative PCA}
Iterative PCA was proposed by \cite{paper:ipca_2008} to estimate the constraint matrix for the generalized case of heteroskedastic errors. It is also capable of identifying the number of linear relations, denoted by \textit{d}, which is not known a priori. The sample covariance matrix of noisy measurements can be expressed as:
\begin{align}
    \mathbf{S_z} = \mathbf{S_x} + \mathbf{\Sigma_e} \label{eq:cov_tmp}
\end{align}
where $\mathbf{S_x} = \frac{1}{N}\mathbf{X}^T \mathbf{X}$ is the covariance matrix of noise-free measurements. The key idea is to scale the noisy measurements matrix with $\mathbf{\Sigma_e}^{-1/2}$, which transforms the problem into an ordinary PCA framework. Let the scaled measurements be $\mathbf{z}_s[k] = \mathbf{\Sigma_e}^{-1/2}\mathbf{z}[k]$. The sample covariance matrix for $N$ samples of scaled measurements denoted by $\mathbf{S_z}_s$ can be expressed as:
\begin{align}
     \mathbf{S_z}_s = \mathbf{S_x}_s + \mathbf{I}_{M \times M}
\end{align}
where $\mathbf{S_x}_s$ denotes the covariance matrix for scaled noise-free measurements and  $\mathbf{I}_{M \times M}$ is identity matrix of dimension of $M \times M$. This relation helps to map the eigenvalues of $\mathbf{S_z}_s$ and $\mathbf{S_x}_s$, which was not the case before scaling in Eq. \ref{eq:cov_tmp}. Consequently, the constraint matrix can also be obtained from the eigenvectors of the scaled covariance matrix with slight modification. 

The estimate of $\mathbf{\Sigma_e}^{-1/2}$ is also obtained from IPCA. The dimension of $\mathbf{A}_0$ is determined by looking at unity eigenvalues rather than zero eigenvalues. For the theoretical basis for the foregoing discussion on the working of IPCA, please refer \cite{paper:ipca_2008}.

This completes the brief discussion on the identification of a linear steady-state model for static systems. In the next subsection, we discuss the extension of PCA and IPCA for dynamic PCA. 
\subsection{Dyanmic PCA \& Dynamic Iterative PCA (DIPCA)}
Dynamic PCA \citep{ku1995disturbance} was one of the earliest extension of PCA for dynamic systems. Consider the class of parametric deterministic SISO linear time-variant dynamic input($u^*$)-output($y^*$) systems described by
\begin{align}
    y^*[k]+ \sum_{i=1}^{n_y} a_i y^*[k-i] = \sum_{j=D}^{n_u} b_j u^*[k-j] 
\end{align}
where $n_y$ and $n_u$ are orders of output and input respectively and \textit{D} is the input-output delay. The EIV identification problem is estimating the coefficients ${\{a_i\}}_{i=1}^{n_y}$, ${\{b_j\}}_{j=D}^{n_u}$ from the measurements of $y^*[k]$ and $u^*[k]$ denoted by $y[k]$ and $u[k]$, respectively. Let the equation order $\eta = \max(n_y, n_u)$.

\cite{ku1995disturbance} proposed to stack the lagged measurements of input and output variables up to order $L$ as shown below:
{\small
\begin{align}
    \mathbf{z}_L[k]& = \begin{bmatrix}
    y[k] & y[k-1] & \ldots y[k-L] & u[k] & u[k-1] & \ldots & u[k-L]
    \end{bmatrix}^T \nonumber \\ 
    \mathbf{Z}_L &= \begin{bmatrix}
    z_L[L] & z_L[L+1] & \ldots & z_L[N]
    \end{bmatrix} \label{eq:Z_stack}
\end{align}
}
The key idea was to apply PCA on the stacked measurements $\mathbf{Z}_L$. The lag order $L$ was determined by trying several lag orders in a sequential manner. As this was a direct extension of PCA, it was limited to handle homoskedastic cases only. 

Dynamic IPCA \citep{paper:dipca_dycops} is proposed to identify the linear dynamic systems in the difference equation form for a SISO system when both input and output measurements are corrupted by errors with different unknown variances. DIPCA uses a two-step procedure to estimate the error variances, order, delay, and model parameters.
\begin{enumerate}
    \item In the first step, the error variances and order of the system are estimated by applying IPCA on a stacked vector of input and output measurements. 
    \item In the second step, the estimated order is used to obtain an appropriately stacked vector (stacking lag is estimated system order). The stacked measurements are scaled using the estimated error variances, and dynamic PCA is used to obtain the model parameters from the eigenvector corresponding to the smallest eigenvalue.
\end{enumerate}
We illustrate the procedure using a simple example of a second order SISO system.
\begin{align}
    y^*[k] + 0.4y^*[k-1] + 0.6y^*[k-2] = 1.2u^*[k-1]
\end{align}
Assuming the availability of $N$ measurements of $y[k]$ and $u[k]$ are available, and the order of system is unknown. The variances of errors added to input and output are unequal, $\sigma^2_{e_y} = 0.24$ and $\sigma^2_{e_u} = 0.1$ (such that SNR is 10). 

The first step of DIPCA is to stack the lagged measurements, as shown in Eq. \eqref{eq:Z_stack}. We construct the data matrix for $L = 3$. The eigenvalues of the scaled data matrix, $\mathbf{S_{z}}_s$ are 
\begin{align}
    \Lambda = \begin{bmatrix}
    24.8 & 21.4 & 12.7 & 10.4 & 9.2& 2.9 & 1.0004 & 0.9996
    \end{bmatrix}
\end{align}
There are two unity eigenvalues indicating the presence of two linear constraints (\textit{d}) among the variables. The estimated noise variances are $\hat{\sigma}^2_{e_y} = 0.2314$ and $\hat{\sigma}^2_{e_u} = 0.0924$ which are almost close to true noise variances.

Now using the number of unity eigenvalues, the order of system can be estimated as:
\begin{align}
    \eta = L - d + 1 = 3 - 2 + 1 = 2
\end{align}
In the second step, the data matrix is reconstructed using a stacking order of $ L= 2$ (estimated order of the system). The data matrix is scaled with $\mathbf{\Sigma_e}^{-1/2}$, where 
\begin{align}
    \mathbf{\Sigma_e} = \hat{\sigma^2}_{e_y} \mathbf{I}_{L+1} \oplus \hat{\sigma^2}_{e_u} \mathbf{I}_{L+1}
\end{align}{}
where $\oplus$ denotes direct sum of two matrices. For instance,
\begin{align}
    \left[ {\begin{array}{cc}a & b \\c & d\\ \end{array} } \right] \oplus \left[ {\begin{array}{cc}e & f \\g & h\\ \end{array} } \right] = \left[ {\begin{array}{cccc}a&b&0&0\\c&d&0&0\\0&0&e & f \\0&0&g & h\\ \end{array} } \right]
\end{align}
The eigenvector corresponding to minimum eigenvalue provides the estimate of the model relating the output and input variables.
\begin{align}
    \mathbf{\hat{A}} = \begin{bmatrix}
    1 & 0.4032 & 0.6004 & 0.0055 & 1.2116 & 0.0083
    \end{bmatrix}
\end{align}
Please note that the coefficients of $u[k]$ and $u[k-2]$ are estimated to be very small as $0.0055$ and $0.0083$, respectively. It can be deduced as insignificant and can be removed by estimating confidence interval using sampling methods such as Monte Carlo methods and boot-strapping. This is illustrated later in the simulation studies section. The identified system in transfer function form is
$$
y[k] = \frac{1.2116q^{-1}}{1+0.4032q^{-1}+0.6004q^{-2}}u[k]
$$
In the next section, we further discuss the extension of the DIPCA algorithm from the SISO system to MISO systems. The algorithm is further modified to estimate the transfer function in minimal realization form. 

\section{Estimating the transfer functions of individual channels in a MISO system}
\label{sec:prop_algo}
We use a simple case study to demonstrate the shortcomings of existing methods and further propose its remedies. Consider a simple second-order MISO system with two inputs as follows:
\begin{align}
    y^*[k] = \frac{1.3q^{-1}}{1-0.2q^{-1}}u_{1}^{*}[k] + \frac{0.7q^{-1}}{1-0.9q^{-1}}u_{2}^{*}[k] \label{eq:true_miso}
\end{align}
We assume the availability of  $N$ measurements of $\{\mathbf{y}[k]\}$, $\{\mathbf{u_1}[k]\}$ and $\{\mathbf{u_2}[k]\}$. Gaussian white noise of zero mean and variance of  $\sigma^2_{e_y} = 2.6868$, $\sigma^2_{e_{u_1}} = 0.9$ and $\sigma^2_{e_{u_2}} = 0.4$ is added to the variables respectively (such that SNR is 10). The difference equation for the above system is as follows:
{\small 
\begin{align}
    y^*[k] - 1.1y^*[k-1] +& 0.18y^*[k-2] = 1.3u_1^*[k-1] -1.17u_1^*[k-2] \nonumber   \\ & +0.7u_2^*[k-1]-0.14u_2^*[k-2] 
\end{align}} 
Applying DIPCA algorithm produces the following difference equation estimate:
{\footnotesize 
\begin{align}
    y^{*}[k] -& 1.0952y^*[k-1] + 0.1789y^*[k-2] = 0.0106u_1^*[k] \nonumber \\ &+1.3150u_1^*[k-1]  
    -1.1738u_1^*[k-2]+0.0098u_2^*[k] \nonumber \\ &+0.7099u_2^*[k-1]-0.1355u_2^*[k-2] \label{eq:dipca_nai}
\end{align}} 
The estimated error variances are:
{\small 
\begin{align}
    \hat{\sigma}^2_{e_y} = 2.6341, \quad \hat{\sigma}^2_{e_{u_1}} = 0.9596, \quad \hat{\sigma}^2_{e_{u_2}} = 0.4206
\end{align}}
which are very close to true variances added. Note that the coefficients, $-0.0106$ and $0.0098$ in Eq. \eqref{eq:dipca_nai} are negligible and can be detected as insignificant by boot-strapping methods. The above difference equation gives us the following transfer functions.
{\small 
\begin{align}
    \hat{G_1}(q^{-1}) &= \frac{1.3150q^{-1}-1.1738q^{-2}}{1-1.0952q^{-1}+0.1789q^{-2}} \nonumber \\
    \hat{G_2}(q^{-1}) &= \frac{0.7099q^{-1}-0.1355q^{-2}}{1-1.0952q^{-1}+0.1789q^{-2}}
\end{align}}
Further the above transfer functions can be factorized as,
{\small 
\begin{align}
    \hat{G_1}(q^{-1}) = \frac{1.3150q^{-1}(1-0.8926q^{-1})}{(1-0.1997q^{-1})(1-0.8954q^{-1})} \nonumber \\
    \hat{G_2}(q^{-1}) = \frac{0.7099q^{-1}(1-0.1908q^{-1}}{(1-0.1997q^{-1})(1-0.8954q^{-1})} \label{eq:dipca_est}
\end{align}}
On careful inspection of $\hat{G_1}(q^{-1})$ and $\hat{G_2}(q^{-1})$ in Eq. \eqref{eq:dipca_est} with $G_1^{true}(q^{-1})$ and $G_2^{true}(q^{-1})$ in \eqref{eq:true_miso}, it can be observed that the estimated transfer function is of higher realization and there exists a pole and zero which are very close to each other numerically. In the next section a systematic procedure is presented for estimating the transfer function in its minimum realization form.

\subsection{Estimating transfer function in minimum realization}
The key idea of the proposed algorithm is to estimate the individual transfer functions for each input rather than the overall transfer function. For this purpose, we decompose the output as the sum of the response due to each input. 

We define $y_1^*[k]$ and $y_2^*[k]$ as follows:
\begin{align}
    y_1^*[k] &= G_1(q^{-1})u_1^*[k] = y^*[k] - y_2^*[k] \nonumber \\ 
    y_2^*[k] & = G_2(q^{-1})u_2^*[k] = y^*[k] - y_1^*[k] \label{eq:y1_star}
\end{align}
where $y_1^*[k]$ is output $y^*[k]$ conditioned on input $u_1^*[k]$ and $y_2^*[k]$ is $y^*[k]$ conditioned on $u_2^*[k]$. The above equation can also be expressed as:
\begin{align}
    y_1^*[k] =  y^*[k] - G_2(q^{-1})u_2^*[k] 
\end{align} 
Let $y_1[k]$ denote the pseudo measurement of $y_1^*[k]$ and from the above equation, it can be expressed as
\begin{align}
   y_1[k] =  y[k] - G_2(q^{-1})u_2[k]  \label{eq:y_1kcomp}
\end{align}
In the above equation  $y[k]$ and $u_2[k]$ are measured and an initial estimate of $G_2(q^{-1})$ is obtained from DIPCA algorithm as shown in the previous subsection. Hence $y_1[k]$ can be computed using Eq. \eqref{eq:y_1kcomp}. The error in the variable $y_1[k]$ is defined as:
\begin{align}
    e_{y_1}[k] = e_y[k] - G_2(q^{-1})e_{u_2}[k]  \label{eq:noise_y1}
\end{align}
This was done to compute the transfer function separately from pseudo measurement $y_1[k]$ and input $u_1[k]$. One of the important aspects of the DIPCA algorithm is the requirement of the noise covariance matrix of pseudo measurement $y_1[k]$. 

One important aspect is that the segregation of output makes the error in pseudo measurement to be colored. We denote the error co-variance of $y_1[k]$ as $\mathbf{R}_{e_{y_1}}$. It contains off diagonal elements whose values are auto co-variance (ACVF) of $e_{y_1}$ at lags 1,2 etc. Using DIPCA we have already estimated the noise variances in the variables $y[k]$, $u_1[k]$ and $u_2[k]$. Further we compute the ACVF of $e_{y_1}[k]$ using these estimated variances. 

Also note that while computing $y_1[k]$ we have only accounted for errors in $y[k]$ and $u_1[k]$ and assumed that the estimated $G_2(q^{-1})$ is true. The key idea is to model $y_1[k]$ and $u_1[k]$ as a SISO system and identify $G_1(q^{-1})$. The procedure to compute ACVF of $e_{y_1}[k]$ from $\hat{\sigma}^2_{e_y}$, $\hat{\sigma}^2_{e_{u_1}}$, $\hat{\sigma}^2_{e_{u_2}}$ and $G_2(q^{-1})$ is described in next section. 
\subsection{Computing the ACVF of $e_{y_1}$[k]}
Recall Wiener-Khinchin theorem \citep{book:arunsir_book} states that, any stationary process with ACVF $\sigma_{vv}[l]$ satisfying 
\begin{align}
    \sum_{l=-\infty}^{\infty}|\sigma_{vv}[l]| < \infty
\end{align}
has the following spectral representation 
\begin{align}
    \sigma_{vv}[l] = \int_{-\pi}^{\pi} \gamma_{vv}(\omega)e^{j\omega l} d\omega
\end{align}
where the stationary process is defined as:
\begin{align}
    v[k] = H(q^{-1})e[k]
\end{align}
and $H(q^{-1})$ is the transfer function relating $v[k]$ and $e[k]$
\begin{align}
    \gamma_{vv}(\omega) = |H(e^{-j\omega})|^2\gamma_{ee}(\omega) = |H(e^{-j\omega})|^2\frac{\sigma^2_{e}}{2\pi} \label{eq:weiner}
\end{align}
Note that the ACVF is symmetric, $\sigma_{vv}[l] = \sigma_{vv}[-l]$. Using the above equations, ACVF of $e_{y_1}[k]$ can be estimated. Let the variables in  Eq \eqref{eq:noise_y1} be denoted as:
\begin{align}
    e_{y_1}[k] = v_y[k] - v_{u_2}[k]
\end{align}
where $v_y[k] = e_y[k]$ and $v_{u_2}[k] = G_2(q^{-1})e_{u_2}[k]$. As  $v_y[k]$ and $v_{u_2}[k]$ are uncorrelated, we derive the following
\begin{align}
    \sigma_{e_{y_1}e_{y_1}}[l] = \sigma_{v_{y}v_{y}}[l] + \sigma_{v_{u_2}v_{u_2}}[l]
\label{eq:acvf_y1}
\end{align}
From Eq. \eqref{eq:weiner}, ACVF of the terms in above equation can be computed as:
\begin{align}
    \sigma_{v_{y}v_{y}}[l] =  \int_{-\pi}^{\pi} \frac{\sigma^2_{e_y}}{2\pi} e^{j\omega l} d\omega = \int_{-\pi}^{\pi} \frac{1}{2\pi} e^{j\omega l} d\omega  \nonumber \\
    \sigma_{v_{u_2}v_{u_2}}[l] = \int_{-\pi}^{\pi} |G_2(e^{-j\omega})|^2\frac{\sigma^2_{e_{u_2}e_{u_2}}}{2\pi}e^{j\omega l} d\omega
\end{align}
Let the ACVF of $e_{y_1}[k]$ at lag $l$ be denoted by $\sigma[l]$ by dropping the subscript for ease of notation. This can be computed using Eq. \eqref{eq:acvf_y1}. For L = 3,
$$
\mathbf{R}_{e_{y_1}}=
  \left[ {\begin{array}{cccc}
   \sigma[0] & \sigma[1] & \sigma[2] & \sigma[3]\\
   \sigma[-1] & \sigma[0] & \sigma[1] & \sigma[2]\\
   \sigma[-2] & \sigma[-1] & \sigma[0] & \sigma[1]\\
   \sigma[-3] & \sigma[-2] & \sigma[-1] & \sigma[0]\\
  \end{array} } \right]  
$$
Let the error-covariance matrix of stacked measurements of pseudo measurement of $y_1[k]$ and input $u_1[k]$ up to lag order L be denoted by $\mathbf{\Sigma}_{e_{y_1}}$. This is required in the DIPCA algorithm while estimating the individual transfer function. It can be estimated using 
\begin{align}
    \mathbf{\Sigma}_{e_{y_1}} = \mathbf{R}_{e_{y_1}} \oplus \hat{\sigma}^2_{u_1} \mathbf{I_{L+1}} \label{eq:cov_ey1}
\end{align}
where $\mathbf{I_{L+1}}$ denotes the identity matrix. This completes 
\subsection{Estimating the transfer functions}
We have discussed the procedure to compute $y_1[k]$ from Eq. \eqref{eq:y1_star} and the error co-variance matrix from Eq. \eqref{eq:cov_ey1}. The problem is simplified to SISO identification with $y_1[k]$ as output and $u_1[k]$ as input. We now perform modified iterative PCA on the data matrix constructed as shown in Eq. \eqref{eq:Z_stack}, which involves two steps.
\begin{itemize}
    \item The first step is choosing a stacking lag L, and scaling the data matrix with $\mathbf{\Sigma}_{e_{y_1}}^{-1/2}$. The order is then computed from $\eta = L - \mathit{d} + 1$, where $\mathit{d}$ - represents the number of unity eigenvalues.
    \item After order estimation, the data matrix is re-stacked up to the estimated order. The stacked measurements are scaled using $\mathbf{\Sigma}_{e_{y_1}}^{-1/2}$, and dynamic PCA is used to obtain the model parameters from the eigenvector corresponding to the minimum eigenvalue.
\end{itemize}
We apply the proposed idea to $y_1[k]$ and $u_1[k]$ and identify the transfer function $G_1(q^{-1})$. $y_1[k]$ is computed using:
$$
y_1[k] =  y[k] - \frac{0.7099q^{-1}-0.1355q^{-2}}{1-1.0952q^{-1}+0.1789q^{-2}}u_2[k]
$$
We compute $\mathbf{\Sigma}_{e_{y_1}}$ using Eq. \eqref{eq:cov_ey1} with L = 2
\footnotesize{$$
\mathbf{R}_{e_{y_1}}=
  \left[ {\begin{array}{ccc}
   \sigma[0] & \sigma[1] & \sigma[2]\\
   \sigma[-1] & \sigma[0] & \sigma[1]\\
   \sigma[-2] & \sigma[-1] & \sigma[0]\\
\end{array} } \right]=
\left[ {\begin{array}{ccc}
   2.3384 & 1.1957 & 1.0839\\
   1.1957 & 2.3384 & 1.1957\\
   1.0839 & 1.1957 & 2.3384\\
\end{array} } \right]
$$}
The estimated error variance of input, denoted by $\hat{\sigma}^2_{e_{u_1}}$ is 0.9596.
\footnotesize{$$
\mathbf{\Sigma}_{e_{y_1}}=
  \left[ {\begin{array}{cccccc}
   2.3384 & 1.1957 & 1.0839 & 0 & 0 & 0\\
   1.1957 & 2.3384 & 1.1957 & 0 & 0 & 0\\
   1.0839 & 1.1957 & 2.3384 & 0 & 0 & 0\\
   0 & 0 & 0 & 0.9596 & 0 & 0\\
   0 & 0 & 0 & 0 & 0.9596 & 0\\
   0 & 0 & 0 & 0 & 0 & 0.9596\\
  \end{array} } \right] 
$$}
\normalsize
The obtained eigenvalues are 
$$
[22.65\hspace{0.2cm}18.1\hspace{0.2cm}11.49\hspace{0.2cm}10.37\hspace{0.2cm}1.0364\hspace{0.2cm}1.1978]
$$
Two unity eigenvalues are identified, and order of the system is $\eta =L - \mathit{d} + 1 = 2 - 2 + 1 = 1$. Now we re-stack the data up to lag 1 and scale it with $\mathbf{\Sigma_{e_{y_1}}}^{-1/2}$ as specified in the above equation. The eigenvector corresponding to least eigenvalue will give the model relating $y_1[k]$ and $u_1[k]$.
$$
y_1[k] - 0.1734y_1[k-1] = -0.0021u_1[k] + 1.2870u_1[k-1]
$$
In the above equation the coefficient of $u_1[k]$ is insignificant and can be dropped by performing hypothesis tests. Hence the estimated transfer function is
$$
\left(1-0.1734q^{-1}\right)y_1[k] = 1.2870q^{-1}u_1[k]
$$
and the original data generating process is
$$
\left( 1-0.2q^{-1} \right) y_1^*[k] = 1.3q^{-1}u_1^*[k]
$$

\subsection{The Proposed Algorithm}
\begin{enumerate}
\small{
\item Given the inputs $u_i[k]$ and the output $y[k]$ construct the data matrix \textbf{Z} and apply DIPCA algorithm to it estimate order, error variances of variables and parameters of the overall model.
\item The estimated difference equation is expressed in transfer function form as
$$
y[k] = \sum_{i=1}^{2} G_i(q^{-1})u_i[k]
$$
\item Compute $y_i[k]$ using 
$$
y_i[k] = y -\sum_{r \ne i}^{} G_r(q^{-1})u_r[k]
$$
\item Apply Wiener-Khinchin theorem to compute the ACVF of $e_{y_{i}}$
$$
\sigma_{v_{y}v_{y}}[l] =  \int_{-\pi}^{\pi} \frac{1}{2\pi} e^{j\omega l} d\omega 
$$
$$
\sigma_{v_{u_r}v_{u_r}}[l] = \int_{-\pi}^{\pi} |G_r(e^{-j\omega})|^2\frac{\sigma^2_{e_{u_r}e_{u_r}}}{2\pi}e^{j\omega l} d\omega 
$$
$$
\sigma_{e_{y_i}e_{y_i}}[l] = \sigma_{v_{y}v_{y}}[l] + \sum_{r \ne i}^{}\sigma_{v_{u_r}v_{u_r}}[l]
$$
\item Choose a stacking lag L and construct $\mathbf{\Sigma}_{e_{y_i}}$ using $\mathbf{\Sigma}_{e_{y_i}} = \mathbf{R}_{e_{y_i}} \oplus \hat{\sigma}^2_{u_i} \mathbf{I_{L+1}}$
\item Construct the data matrix and scale it with estimated $\mathbf{\Sigma}_{e_{y_i}}^{-1/2}$ to estimate the order of the system using $\eta = L-\mathit{d}+1$
\item Re-stack the measurements up to the estimated order and obtain the eigenvector corresponding to least eigenvalue to identify the transfer function relating $y_i[k]$ and $u_i[k]$ which is $G_i(q^{-1})$ 
}
\end{enumerate}
\normalsize
The next section presents  results from simulations studies to demonstrate effectiveness of the proposed algorithm.

\section{SIMULATION RESULTS}
\label{sec:siml_stud}
The case study pertains to the second-order (overall) two input single output system in Eq. \eqref{eq:true_miso}. We generate  $5000$ samples of each variable follwing the difference equation and corrupt them with unequal error variances of $\sigma^2_y = 2.68$, $\sigma^2_{u_1} = 0.9$ and $\sigma^2_{u_2} = 0.4$. (SNR = 10). A snapshot of few samples is shown below:
\begin{figure}[thpb]
  \centering
  \includegraphics[scale=0.4]{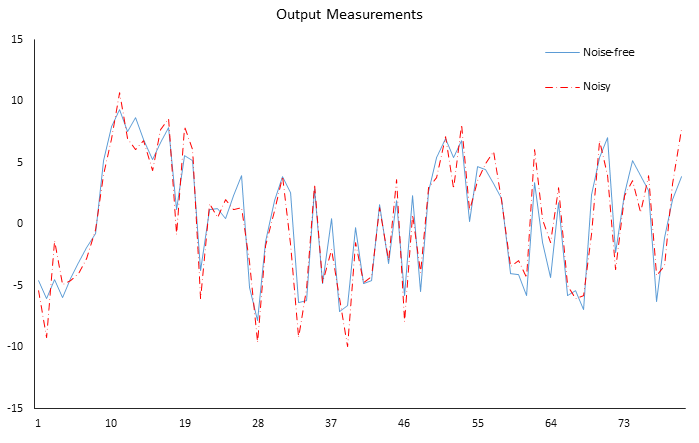}
  \caption{Output measurement of a realization}
  \label{figurelabel1}
\end{figure}
\begin{figure}[thpb]
  \centering
  \includegraphics[scale=0.40]{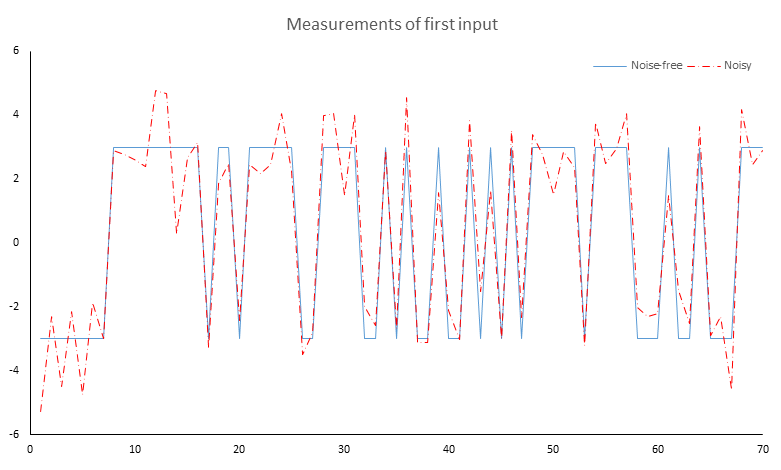}
  \caption{Measurements of $u_1$ for a realization}
  \label{figurelabel3}
\end{figure}
\begin{figure}[thpb]
  \centering
  \includegraphics[scale=0.4]{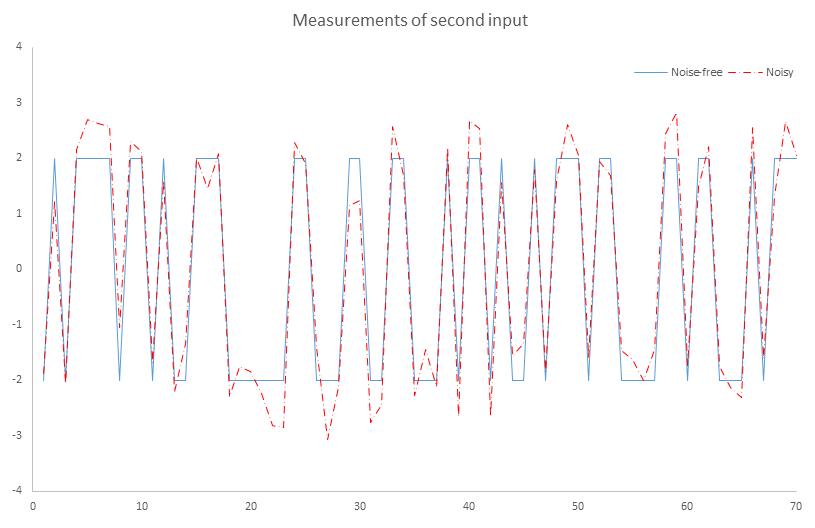}
  \caption{Measurements of $u_2$ for a realization}
  \label{figurelabel4}
\end{figure}

The proposed algorithm is applied with stacking lag, $L = 5$. Four near unity eigenvalues were observed as seen from the last five of eighteen eigenvalues reported below:
\begin{align}
    \Lambda_{18 \times 18} = [2.9\hspace{0.2cm}1.02\hspace{0.2cm}1.008\hspace{0.2cm}0.99\hspace{0.2cm} 0.9733]
\end{align}
\begin{figure}[thpb]
  \centering
  \includegraphics[scale=0.45]{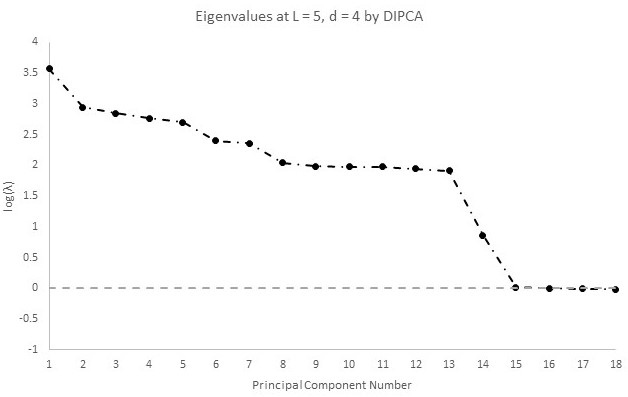}
  \caption{The eigenvalues obtained for L = 5}
  \label{figurelabel2}
\end{figure}
Further, the estimate of noise variance \\ $\hat{\Sigma}_e = \text{diag}(\begin{bmatrix} 2.7461 & 0.8580 & 0.4989 \end{bmatrix})$ closely matches with the true value used in simulations. The order of the system is estimated to be $n_y = 5 - 4 + 1 = 2$, thus identifying the order correctly. The measurements are now re-stacked up to order and scaled with estimated error variance matrix yielding the following difference equation
\small{
$$
y[k] = 1.1129y[k-1] - 0.1855y[k-2] + 0.0028u_1[k] + 1.3815u_1[k-1] \linebreak
$$
$$
 - 1.1835u_1[k-2] + 0.0020u_2[k] + 0.6548u_2[k-1]-0.1611u_2[k-2]
$$
}
\normalsize
The insignificant coefficients can be removed by hypothesis testing of estimated coefficients and hence the estimated difference equation can be expressed as 
\small{
$$
y[k] = G_1(q^{-1})u_1[k] + G_2(q^{-1})u_2[k]
$$where,
$$
\hat{G_1}(q^{-1}) = \frac{1.3851q^{-1}-1.1835q^{-2}}{1-1.1129q^{-1}+0.1855q^{-2}}
$$
$$
\hat{G_2}(q^{-1}) = \frac{0.6548q^{-1}-0.1611q^{-2}}{1-1.1129q^{-1}+0.1855q^{-2}}
$$
}
\normalsize
The next step is to run the proposed algorithm on each input separately as discussed in the previous section. We observe 2 unity eigenvalues for a chosen L = 2, that are reported below
\footnotesize{
$$
\Lambda_{6 \times 6} = \text{diag}([29.3041 \hspace{0.2cm} 21.7838 \hspace{0.2cm}17.8950 \hspace{0.2cm}6.2916 \hspace{0.2cm}1.1957 \hspace{0.2cm} 0.9941]) 
$$}
\normalsize
The order of the transfer function is $\eta = 2 - 2 + 1 = 1$. The next step is to re-stack the measurements up to the estimated order and utilize the eigenvector corresponding to the least eigenvalue to compute the difference equation 
$$
y[k] - 0.1510y[k-1] = 0.0017u_1[k] + 1.2368u_1[k-1]
$$
The same procedure can be applied to other input. The next step is to check the significance level of the estimated parameters. Monte-Carlo simulations of 100 runs are performed for this purpose, and the results are mentioned in Table 1. 
\begin{table}[ht]
\footnotesize{\caption{The parameter estimates of the transfer functions in Eq. \eqref{eq:true_miso} }}
\normalsize
\begin{center}
    \begin{tabular}{|c c c c|} 
        \hline
            Parameter & True Value & mean & std. dev \\ [0.5ex] 
        \hline\hline
            a$_1$ & -0.2 & -0.1924 & 0.0320 \\
        \hline
            b$_0$ & 0 & 0.0006 & 0.0120 \\
        \hline
            b$_1$ & 1.3 & 1.3154 & 0.0898 \\
        \hline
            c$_1$ & -0.9 & -0.8782 & 0.0205 \\
        \hline
            d$_0$ & 0 & 0.0056 & 0.0262 \\
        \hline
            d$_1$ & 0.7 & 0.7021 & 0.0901 \\
        \hline
    \end{tabular}
\end{center}
\end{table}
Here, the transfer functions are represented as
\scriptsize{
$$
G_1(q^{-1}) = \frac{b_{0}+b_{1}q^{-1}}{1+a_{1}q^{-1}}, G_2(q^{-1}) = \frac{d_{0}+d_{1}q^{-1}}{1+c_{1}q^{-1}}
$$
}

\begin{table}[ht]
\caption{The noise variance estimates of variables for system in Eq. \eqref{eq:true_miso}}
\normalsize
    \begin{center}
    \begin{tabular}{|c c c c|}
            \hline
                Para. & True value & mean & std. dev \\ [0.5ex] 
            \hline
                $\sigma_{\epsilon_y}^2$ & 2.6868 & 2.6343 & 0.5743\\
            \hline
                $\sigma_{\epsilon_{u_1}}^2$ & 0.9000 & 0.9023 & 0.0434\\
            \hline
                $\sigma_{\epsilon_{u_2}}^2$ & 0.4000 & 0.3752 & 0.0259\\
            \hline
        \end{tabular}
    \end{center}
\end{table}
\normalsize
The actual value used in data generating process is presented in  Eq. \eqref{eq:true_miso}. It can be clearly observed that all the  estimates are close to the true value. 
\section{CONCLUSIONS}
\label{sec:conc}
In this paper, we have presented a systematic method for recovering the transfer functions in their minimum realization form of a linear multi-input single-output system from the measurements of inputs and outputs in the EIV case. Using the difference equation estimated from DIPCA algorithm, we proposed a method for determining the minimum realization of the transfer functions. Simulation studies show that the proposed method is efficient for identifying the transfer functions purely from data with minimal user intervention.

\bibliographystyle{ieeetr}
\bibliography{ifacconf}             
                                                   







\end{document}